\documentstyle[epsfig,amsmath,12pt]{article}
\author{Petr Z\'avada\\
\\\
{\it Institute of Physics, Academy of Sciences of Czech Republic}\\
{\it Na Slovance 2, CZ-180 40 Prague 8}\\
E-mail: zavada@fzu.cz}
\title{The structure functions and parton momenta distribution in the hadron  rest system
}
\begin{document}

\maketitle
\begin{abstract}
The alternative to the standard formulation of QPM in the infinite momentum
frame is suggested. The proposed approach does not require any extra
assumptions in addition, consistently takes into account the parton
transversal momenta and does not prefer any special reference system. The
standard approach is involved as a limiting case. In the result the modified
relations between the structure and distribution functions are obtained
together with some constraint on their shape. The comparison with
experimental data offers a speculation about values of effective masses of
quarks, which emerge as a free parameter in the approach.
\end{abstract}

\renewcommand{\theequation} {\thesection .\arabic{equation}}

\section{Introduction}

The deep inelastic scattering (DIS) of leptons on the nucleons and nuclei
has been since early seventies the powerful tool for investigation of the
nucleon internal structure and simultaneously has served as an crucial test
of the related theory - QCD. For recent results in this field see e.g. \cite
{rev} and citations therein.

The quark-parton model (QPM), motivated by the experimental data, is
extraordinarily simple if formulated in the reference system in which the
nucleon is fast moving (infinite momentum frame - IMF). Namely in this
system the Bjorken scaling variable $x_B$ can be approximately identified
with the momentum fraction of the nucleon carried by a parton and
experimentally measured structure functions can be easily related to the
combinations of distribution functions expressed in terms of $x_B$. The
distribution functions extracted from the experimental data by the global
analysis (see e.g. \cite{parval}) relying on QPM+QCD represent basic
elements of the present picture of nucleons and other hadrons.

In this paper we attempt to cope with the not only aesthetic drawback of QPM
which in the standard formulation has a good sense only in the {\it preferred%
} reference system - IMF. The idea of alternatives to the QPM postulated in
IMF is not new, the possibility to obtain in some approximation the
structure and distribution functions from a definite parton model formulated
in the nucleon rest frame has been shown e.g. in \cite{fra1},\cite{fra2},%
\cite{cle} and recently \cite{fra},\cite{bha}. We suggest rather a
consistent modification of the general standard formulation which does not
adhere necessarily to IMF and simultaneously does not require any special
assumptions in addition. The basis of our considerations are only kinematics
and mathematics.

The paper is organized as follows. In the following section the basic
kinematic quantities related to the DIS are introduced and particularly
the meaning of variable $x_B$ is discussed. In the Sec.3 we formally apply
the standard assumptions of the QPM to the nucleon in its rest system (LAB)
and compare the results with those normally related to the IMF. The Sec.4 is
devoted to the discussion from more physical point of view together with the
glance at experimental data on proton structure function $F_2$. The last
section shortly summarizes the possible conclusions.

\section{Kinematics}

\setcounter{equation}{0}

\begin{figure}
\begin{center}
\epsfig{file=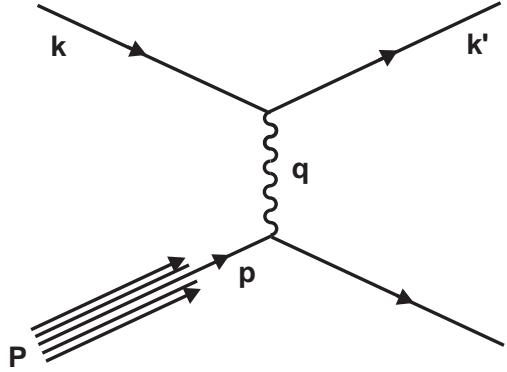, height=5cm}
\end{center}
\caption {Diagram describing DIS as a one photon exchange between the 
charged lepton and parton.\label{fg1}}
\end{figure}

First of all let us recall some basic notions used in the description of DIS
and the interpretation of the experimental data on the basis of QPM. The
process is usually described (see Fig.\ref{fg1}) by the variables 
\begin{equation}
\label{k1}q^2\equiv -Q^2=(k-k^{\prime })^2,\qquad x_B=\frac{Q^2}{2Pq} 
\end{equation}
As a rule, lepton mass is neglected, i.e. $k^2=k^{\prime 2}=0$. Important
assumption of QPM is that struck parton remains on-shell, that implies

\begin{equation}
\label{k2}q^2+2pq=0 
\end{equation}
Bjorken scaling variable $x_B$ can be interpreted as the fraction of the
nucleon momentum carried by the parton in the nucleon infinite momentum
frame (IMF). The motivation of this statement can be explained as follows.
Let us denote

\begin{equation}
\label{k3}p(lab)\equiv (p_0,p_1,p_2,p_3),\qquad P(lab)\equiv
(M,0,0,0),\qquad q(lab)\equiv (q_0,q_1,q_2,q_3) 
\end{equation}
fourmometa of the parton, nucleon and exchanged photon in the nucleon rest
system (LAB). The Lorentz boost to the IMF (in the direction of collision
axis) gives

\begin{equation}
\label{k4}p(\inf )\equiv (p_0^{\prime },p_1^{\prime },p_2,p_3),\qquad P(\inf
)\equiv (P_0^{\prime },P_1^{\prime },0,0),\qquad q(\inf )\equiv (q_0^{\prime
},q_1^{\prime },q_2,q_3) 
\end{equation}
where for $\beta \rightarrow -1$

\begin{equation}
\label{k5}p_0^{\prime }=p_1^{\prime }=\gamma (p_0+p_1),\quad P_0^{\prime
}=P_1^{\prime }=\gamma M,\qquad \gamma =1/\sqrt{1-\beta ^2} 
\end{equation}
If we denote

\begin{equation}
\label{k6}x\equiv \frac{p_0^{\prime }}{P_0^{\prime }}=\frac{p_1^{\prime }}{%
P_1^{\prime }}=\frac{p_0+p_1}M 
\end{equation}
then one can write

\begin{equation}
\label{k7}p(\inf )=xP(\inf )+(0,0,p_2,p_3) 
\end{equation}
Now let the lepton has initial momentum $k(lab)\equiv (k_0,-k_0,0,0)$. If we
denote $\nu \equiv k_0-k_0^{\prime }$ and $q_L\equiv q_1,$ then $q_L<0$ and
from Eqs. (\ref{k1}), (\ref{k2}) it follows

\begin{equation}
\label{k8}x_B=\frac{pq}{Pq}=\frac{p_0\nu +\mid q_L\mid p_1}{M\nu }-\frac{
\overrightarrow{p_T}\overrightarrow{q_T}}{M\nu } 
\end{equation}
where $\overrightarrow{p_T},\ \,\overrightarrow{q_T}$ are the parton and
photon transversal momenta . Obviously

$$
k^{\prime 2}=(k-q)^2=k^2+q^2-2k_0\nu +2k_0\mid q_L\mid =0 
$$

\begin{equation}
\label{ka10}\frac{\mid q_L\mid }\nu =1+\frac{Q^2}{2k_0\nu }=1+\frac
M{k_0}x_B 
\end{equation}
Using this relation the Eq.(\ref{k8}) can be modified

\begin{equation}
\label{ka8}x_B=\frac{p_0+p_1}M+\frac{p_1}{k_0}x_B-\frac{\overrightarrow{p_T} 
\overrightarrow{q_T}}{M\nu } 
\end{equation}
therefore if the lepton energy is sufficiently high, so $p_1/k_0\approx 0$,
one can write

\begin{equation}
\label{k9}x_B=x-\frac{p_T\ \,q_T}{M\nu }\cos \varphi 
\end{equation}
where $\varphi $ is the angle between the parton and photon momenta in the
transversal plane.

So, if parton transversal momenta are neglected, $x_B$ really represents
fraction of momentum (\ref{k6}). In a higher approximation the
experimentally measured $x_B$ being an integral over $\varphi $ is
effectively smeared with respect to the fraction $x-$ which is not
correlated with $\varphi .$ An estimation of the second term in the last
equation can be done as follows. Because

$$
q^2=\nu ^2-\left| \overrightarrow{q}\right| ^2 
$$

\begin{equation}
\label{kb10}\left( \frac{\overrightarrow{q}}\nu \right) ^2=1+\frac{Q^2}{\nu
^2}=1+\frac{4M^2}{Q^2}x_B^2 
\end{equation}
then (\ref{ka10}), (\ref{kb10}) give

\begin{equation}
\label{kba10}\frac{q_T}\nu =\sqrt{\left( \frac{\overrightarrow{q}}\nu
\right) ^2-\left( \frac{q_L}\nu \right) ^2}=\sqrt{\left( \frac{4M^2}{Q^2}- 
\frac{M^2}{k_0^2}\right) x_B^2-\frac{2M}{k_0}x_B}<\frac{2Mx_B}{\sqrt{Q^2}} 
\end{equation}
therefore for $M/k_0\approx 0$ we obtain

\begin{equation}
\label{kd10}\frac{pq}{M\nu }=\frac{p_0+p_1}M-\frac{2p_Tx_B}{\sqrt{Q^2}}\cos
\varphi 
\end{equation}
and

\begin{equation}
\label{kc10}x_B=x-\frac{2p_Tx_B}{\sqrt{Q^2}}\cos \varphi 
\end{equation}
Therefore $x_B$ can be at sufficiently high $Q^2$ considered as a good
approximation of $x$ (and vice versa), on the end of next section we shall
suggest how to treat this correction more accurately.

Let us note, the parameter $x$ (\ref{k6}) can be expressed also as 
\begin{equation}
\label{d1}x\equiv \frac{p_0+p_1}{P_0+P_1} 
\end{equation}
and identified with light cone variable, which can be expressed also in
terms of rapidity and transversal mass

\begin{equation}
\label{da1}x\equiv \frac{m_T}M\exp (y-y_0),\qquad m_T\equiv \sqrt{p_T^2+m^2}%
,\qquad y\equiv \frac 12\ln \ \frac{p_0+p_1}{p_0-p_1} 
\end{equation}
where $y_0$ denotes the proton rapidity. In this form the parameter $x$ is
invariant with respect to any Lorentz boost along the collision axis.

Now, if we assume parton phase space is spherical (in LAB) and rather
idealized scenario in which the parton has a mass $%
m^2=p_0^2-p_1^2-p_2^2-p_3^2$, then further relations can be obtained.

1) {\it variable x}

\noindent
From Eq.(\ref{k6}) and the condition $x\leq 1$ it can be shown

\begin{equation}
\label{k11}x\geq \frac{m^2}{M^2} 
\end{equation}

\begin{equation}
\label{k12}\sqrt{p_1^2+p_2^2+p_3^2}\leq p_m\equiv \frac{M^2-m^2}{2M},\qquad
p_T^2\leq M^2(x-\frac{m^2}{M^2})(1-x) 
\end{equation}
Obviously, the highest value of $p_1$ is reached if $p_T=0$ and 
\begin{equation}
\label{ka12}x=\frac{\sqrt{p_1^2+m^2}+p_1}M=1 
\end{equation}
which gives 
\begin{equation}
\label{kb12}p_{1\max }=p_m\equiv \frac{M^2-m^2}{2M} 
\end{equation}
Then spherical symmetry implies 
\begin{equation}
\label{kc12}\sqrt{p_1^2+p_2^2+p_3^2}\leq p_m 
\end{equation}
i.e. the first relation in (\ref{k12}) is proved. Apparently, the minimal
value of $x$ is reached for $p_1=-p_m$ and $p_T=0$. After inserting to (\ref
{k6}) one gets (\ref{k11}). Finally, the relation (\ref{k6}) implies 
\begin{equation}
\label{kd12}p_1=\frac{M^2x^2-m^2-p_T^2}{2Mx} 
\end{equation}
which inserted to modified relation (\ref{kc12}) 
\begin{equation}
\label{ke12}p_1^2+p_T^2\leq \left( \frac{M^2-m^2}{2M}\right) ^2 
\end{equation}
after some computation gives the second relation in (\ref{k12}).

2) {\it variable} $x_B$

\noindent
Let us express $x_B$ in the LAB

\begin{equation}
\label{k13} x_B=\frac{pq}{Pq}=\frac{p_0\nu -\overrightarrow{p}\ 
\overrightarrow{q}}{M\nu }=\frac 1M\left( \sqrt{m^2+\left| \overrightarrow{p}%
\right| ^2}-\frac{\overrightarrow{p}\ \overrightarrow{q}}\nu \right) 
\end{equation}
and estimate its minimal value. With the use of (\ref{kb10}) we obtain

\begin{equation}
\label{k15}x_B\geq \frac 1M\left( \sqrt{m^2+p_m^2}-p_m\sqrt{1+\frac{4M^2 }{%
Q^2}x_B^2}\right) 
\end{equation}
Since 
\begin{equation}
\label{ka15} \sqrt{1+\frac{4M^2}{Q^2}x_B^2}\leq 1+\frac{2M^2}{Q^2}x_B^2 
\end{equation}
and

\begin{equation}
\label{k16} \frac 1M\left( \sqrt{m^2+p_m^2}-p_m\right) =\frac{m^2}{M^2} 
\end{equation}
relation (\ref{k15}) can be rewritten

\begin{equation}
\label{k17}x_B\geq \frac{m^2}{M^2}-\frac{2M p_m}{Q^2}x_B^2\geq \frac{m^2}{M^2%
}-\frac{2M p_m}{Q^2}\frac{m^4}{M^4}=\frac{m^2}{M^2}(1-\frac{2p_m}M\frac{m^2}{%
Q^2}) 
\end{equation}
i.e. for $m^2\ll Q^2$ lower limit of $x_B$ coincides with the limit (\ref
{k11}).

\section{Distribution of partons in the nucleon rest system}

\setcounter{equation}{0}

In this section we imagine partons as a gas (or a mixture of gases) of quasi
free particles filling up the nucleon volume. The prefix {\it quasi }means
that partons bounded inside the nucleon behave at the interaction with
external photon probing the nucleon as free particles having the fourmomenta
on mass shell. This is standard assumption of QPM, but whereas in the IMF
parton masses are ''hidden'', in the description related to the LAB the
masses will be present.

In the next section we shall discuss in which extent the results obtained
for this idealized picture could be applied for more realistic scenario.

\subsection{Deconvolution of the distribution function}

Let us suppose $F(x)$ is the distribution function of some sort of partons
given in terms of variable $x$ ($\ref{k6})$ and these partons are assumed to
have the mass $m$. If the spherical symmetry is assumed in the hadron rest
system and $G(p_0)d^3p$ is the number of partons in the element of the phase
space, then the distribution function $F(x)$ can be expressed as the
convolution

\begin{equation}
\label{eq4}F(x)=\int \delta \left( \frac{p_0+p_1}M-x\right)
G(p_0)d^3p,\qquad p_0=\sqrt{m^2+p_1^2+p_2^2+p_3^2} 
\end{equation}
Using the set of integral variables $h,p_0,\varphi $ instead of $p_1,p_2,p_3$

\begin{equation}
\label{eq5}p_1=h,\qquad p_2=\sqrt{p_0^2-m^2-h^2}\sin \varphi ,\qquad p_3= 
\sqrt{p_0^2-m^2-h^2}\cos \varphi 
\end{equation}
the integral (\ref{eq4}) can rewritten

\begin{equation}
\label{eq6}F(x)=2\pi \int_m^{E_{\max }}\int_{-H}^{+H}\delta \left( \frac{%
p_0+h}M-x\right) G(p_0)p_0dhdp_0,\qquad H=\sqrt{p_0^2-m^2} 
\end{equation}
First of all we calculate inner integral within limits $\pm H$ depending on $%
p_0.$ For given $x$ and $p_0$ there contributes only $h$ for which

\begin{equation}
\label{eq7}p_0+h=Mx 
\end{equation}
but simultaneously $h$ must be inside the limits

\begin{equation}
\label{eq8}-\sqrt{p_0^2-m^2}\leq h\leq \sqrt{p_0^2-m^2} 
\end{equation}
which means, that for

\begin{equation}
\label{eq9}p_0+\sqrt{p_0^2-m^2}<Mx 
\end{equation}
or equivalently for

\begin{equation}
\label{eq10}p_0<\xi \equiv \frac{Mx}2+\frac{m^2}{2Mx} 
\end{equation}
considered integral gives zero. For $p_0>\xi $, when the both conditions (%
\ref{eq7}), (\ref{eq8}) are compatible for some value $h$ the integral can
be evaluated

\begin{equation}
\label{eq11}\int_{-H}^{+H}\delta \left( \frac{p_0+h}M-x\right)
G(p_0)p_0dh=MG(p_0)p_0 
\end{equation}
Therefore the integral (\ref{eq6}) can be expressed

\begin{equation}
\label{eq12}F(x)=2\pi M\int_\xi ^{E_{\max }}G(p_0)p_0dp_0 
\end{equation}
Let us note, the equation similar to this appears already in \cite{fra1} but
with the structure function $F_2 (x)$ instead of the distribution one. We
shall deal with the $F_2$ in the next subsection, where it will be shown,
that the corresponding equation is more complicated. For a comparison see
also \cite{cle}, where on the place of $G(p_0 )$ the statistical
distribution characterized by some temperature and chemical potential is
used.

Next, from the relation (\ref{eq10}) we can express $x$ as a function $\xi $

\begin{equation}
\label{eq13}x_{\pm }=\frac{\xi \pm \sqrt{\xi ^2-m^2}}M 
\end{equation}
Using the relations (\ref{k11}), (\ref{eq10}) one can easily check

\begin{equation}
\label{eq14}1\geq x_{+}\geq \frac mM\geq x_{-}\geq \frac{m^2}{M^2},\qquad
E_{\max }=\frac{M^2+m^2}{2M}\geq \xi \geq m 
\end{equation}
First let us insert $x_{+}$ into (\ref{eq12})

\begin{equation}
\label{eq16}F\left( \frac{\xi +\sqrt{\xi ^2-m^2}}M\right) =2\pi M\int_\xi
^{E_{\max }}G(p_0)p_0dp_0 
\end{equation}
Differentiation in respect to $\xi $ gives

\begin{equation}
\label{eq17}G(\xi )=-\frac 1{2\pi M^2}F^{\prime }\left( \frac{\xi +\sqrt{\xi
^2-m^2}}M\right) \left( \frac 1\xi +\frac 1{\sqrt{\xi ^2-m^2}}\right) 
\end{equation}
Now we integrate the density $G(p_0)$ over angular variables obtaining

\begin{equation}
\label{eq18}P(p_0)dp_0\equiv \int_\Omega G(p_0)d^3p=4\pi G(p_0)p_0\sqrt{%
p_0^2-m^2}dp_0 
\end{equation}
and after inserting into (\ref{eq17}) we get

\begin{figure}
\begin{center}
\epsfig{file=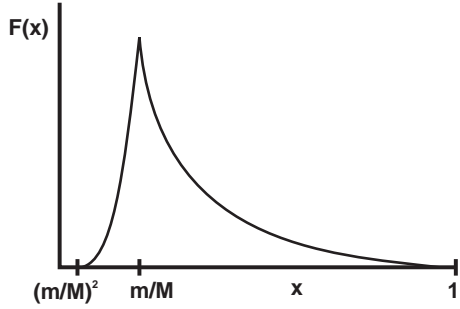, height=5cm}
\end{center}
\caption {Example of the function obeying Eqs.(\ref{eq24}), (\ref{eq24a}).}
\label{fg2}
\end{figure}

$^{}$%
\begin{equation}
\label{eq19}P(p_0)dp_0=-2F^{\prime }\left( \frac{p_0+\sqrt{p_0^2-m^2}}%
M\right) \frac{p_0+\sqrt{p_0^2-m^2}}M\frac{dp_0}M 
\end{equation}
Second root $x_{-}$ gives very similar result

\begin{equation}
\label{eq19a}P(p_0)dp_0=+2F^{\prime }\left( \frac{p_0-\sqrt{p_0^2-m^2}}%
M\right) \frac{p_0-\sqrt{p_0^2-m^2}}M\frac{dp_0}M 
\end{equation}
From the definition

\begin{equation}
\label{eq19b}x_{\pm }\equiv \frac{p_0\pm \sqrt{p_0^2-m^2}}M 
\end{equation}
the useful relations easily follow

\begin{equation}
\label{eq19c}x_{+}x_{-}=\frac{m^2}{M^2},\qquad x_{+}+x_{-}=\frac{2p_0}%
M,\qquad x_{+}-x_{-}=\frac{2\sqrt{p_0^2-m^2}}M 
\end{equation}

\begin{equation}
\label{eq19d}\frac{dp_0}M=\frac 12(1-\frac{m^2}{M^2x_{\pm }^2})dx_{\pm
},\qquad \frac{dx_{+}}{x_{+}}=-\frac{dx_{-}}{x_{-}} 
\end{equation}
Now, the equations (\ref{eq19}), (\ref{eq19a}) can be joined 
\begin{equation}
\label{eq20}P(p_0)=\mp \frac 2MF^{\prime }(x_{\pm })x_{\pm } 
\end{equation}
How to understand the two different partial intervals (\ref{eq14}) of $x$
give independently the complete distribution $P(p_0)$ in Eq.(\ref{eq20})? It
is due to the fact that e.g. $x_{-}$ represents in the integral (\ref{eq4})
the region

\begin{equation}
\label{eq22}\frac{\sqrt{p_1^2+p_T^2+m^2}+p_1}M=x_{-}\leq \frac mM 
\end{equation}
given by the paraboloid

\begin{equation}
\label{eq23}p_T^2\leq 2m\left| p_1\right| ,\qquad p_1\leq 0 
\end{equation}
containing complete information about $G(p_0)$ which is spherically
symmetric. The similar argument is valid for $x_{+}$ representing the rest
of sphere. The Eqs.(\ref{eq19}), (\ref{eq19a}) imply the similarity of $F(x)$
in both intervals

\begin{equation}
\label{eq24}\frac{F^{\prime }(x_{+})x_{+}}{F^{\prime }(x_{-})x_{-}}=-1 
\end{equation}
which with the use of second relation (\ref{eq19d}) can be easily shown to
be equivalent to 
\begin{equation}
\label{eq24a}F(x_{+})=F(x_{-}) 
\end{equation}
The relation (\ref{eq20}) implies the distribution function $F(x)$ should be
increasing for $(m/M)^2<x<m/M$ and decreasing for $m/M<x<1$ e.g. as shown in
Fig.\ref{fg2}. Now let us calculate the following integrals. 

{\it The total number $N$ of partons:}

$$
N=\int_m^{E_{\max }}P(p_0)dp_0=-\int_{m/M}^1F^{\prime }(x_{+})(x_{+}-\frac{%
m^2}{M^2x_{+}})dx_{+}= 
$$

\begin{equation}
\label{eq25}=-\int_{m/M}^1F^{\prime
}(x_{+})x_{+}dx_{+}+\int_{m/M}^1F^{\prime }(x_{+})x_{-}dx_{+} 
\end{equation}
The last integral can be modified with the use of (\ref{eq19d}), (\ref{eq24})

\begin{equation}
\label{eq25a}\int_{m/M}^1F^{\prime
}(x_{+})x_{-}dx_{+}=-\int_{m/M}^1F^{\prime }(x_{-})x_{-}^2\frac{dx_{+}}{x_{+}%
}=\int_{m/M}^{m^2/M^2}F^{\prime }(x_-)x_{-}dx_- 
\end{equation}
Then integration by parts gives

\begin{equation}
\label{eq25b}N=-\int_{m^2/M^2}^1F^{\prime }(x)xdx=\int_{m^2/M^2}^1F(x)dx 
\end{equation}
{\it The total energy $E$ of partons:}

$$
E=\int_m^{E_{\max }}P(p_0)p_0dp_0=-\int_{m/M}^1F^{\prime }(x_{+})(x_{+}- 
\frac{m^2}{M^2x_{+}})\frac M2(x_{+}+x_{-})dx_{+}= 
$$

\begin{equation}
\label{eq26}=-\frac M2\int_{m/M}^1F^{\prime }(x_{+})(x_{+}^2-x_{-}^2)dx_{+} 
\end{equation}
A similar procedure as for $N$ then gives the result

\begin{equation}
\label{eq27}E=-\frac M2\int_{m^2/M^2}^1F^{\prime
}(x)x^2dx=M\int_{m^2/M^2}^1F(x)xdx 
\end{equation}
Therefore, the both descriptions based either on IMF variable $x$ or the
parton energy $p_0$ in the LAB give the consistent results on the total
number of partons and the fraction of energy carried by the partons.

\subsection{The structure function}

An important connection between the structure and distribution functions can
be derived by a few (equivalent) ways, see e.g. textbooks \cite{fey},\cite
{clo},\cite{ait}. In this paper we confine ourself to the electromagnetic
unpolarized structure functions assuming spin 1/2. The general form of cross
section for the scattering {\it electron + proton} and {\it electron + point
like, Dirac particle} can be written 
\begin{equation}
\label{sf00}d\sigma (e^{-}+p)=\frac{e^4}{q^4}\frac 1{4\sqrt{(kP)^2-m_e^2M^2}%
}K^{\alpha \beta }W_{\alpha \beta }4\pi M\frac{d^3k^{\prime }}{2k_0^{\prime
}(2\pi )^3} 
\end{equation}
\begin{equation}
\label{sf01}d\sigma (e^{-}+l)=\frac{e^4}{q^4}\frac 1{4\sqrt{(kp)^2-m_e^2m_l^2%
}}K^{\alpha \beta }L_{\alpha \beta }2\pi \delta ((p+q)^2-m^2)\frac{%
d^3k^{\prime }}{2k_0^{\prime }(2\pi )^3} 
\end{equation}
where electron tensor has the standard form 
\begin{equation}
\label{sf02}K^{\alpha \beta }=2(k^\alpha k^{\prime \beta }+k^{\prime \alpha
}k^\beta +g^{\alpha \beta }\frac{q^2}2) 
\end{equation}
and the remaining hadron and lepton tensors $W_{\alpha \beta },$ $L_{\alpha
\beta }$ can be written in the ''reduced'' shape 
\begin{equation}
\label{sf03}W_{\alpha \beta }=\frac{P_\alpha P_\beta }{M^2}W_2-g_{\alpha
\beta }W_1 
\end{equation}
\begin{equation}
\label{sf04}L_{\alpha \beta }=4p_\alpha p_\beta -2g_{\alpha \beta }pq 
\end{equation}
General assumption that the scattering on proton is realized via scattering
on the partons implies 
\begin{equation}
\label{sf05}d\sigma (e^{-}+p)=\int F(\xi )d\sigma (e^{-}+l)d\xi 
\end{equation}
where $F(\xi )$ is a function describing distribution of partons according
to some parameter(s) $\xi .$ Now, if $F(\xi )$ is substituted by the usual
distribution function and we assume

\begin{equation}
\label{sf2}p_\alpha \approx \xi P_\alpha 
\end{equation}
then it is obvious, that Eq.(\ref{sf05}) will be fulfilled provided that 
\begin{equation}
\label{sf3}P_\alpha P_\beta \frac{W_2}{M^2}-g_{\alpha \beta }W_1=\frac
1M\int \frac{F(\xi )}\xi (2\xi ^2P_\alpha P_\beta -g_{\alpha \beta }\,\xi
Pq)\delta ((\xi P+q)^2-m^2)d\xi 
\end{equation}
For simplicity in this equation and anywhere in the next the weighting by
the parton charges is omitted. In fact the Eq.(\ref{sf3}) is just a master
equation in \cite{fey}(lesson 27, Eq.(27.4)), from which the known relations
are derived 
\begin{equation}
\label{sf4}2MW_1(q^2,\nu )=\frac{F_2(x)}x,\qquad xF(x)=F_2(x)\equiv \nu
W_2(q^2,\nu ),\qquad x\equiv \frac{-q^2}{2M\nu } 
\end{equation}
Here, let us point out, this result is based on the approximation (\ref{sf2}%
), which is acceptable in IMF, but in addition only if {\it parton
transversal momenta are neglected}. Actually, relation (\ref{sf2}) would be
exact only in the (unrealistic) case, when the partons are without any
motion inside the nucleon, then the distribution function describes momenta
fractions in {\it any} reference frame (including IMF), therefore describes
also distribution of parton masses $\xi =m/M$.

Before repeating the above procedure for our distribution $G(p_0)d^3p$ in
LAB, one has correctly account for the flux factor corresponding to partons
moving inside the proton volume. For $k\equiv (k_0,-k_0,0,0)$ the flux
factor in (\ref{sf01}) 
\begin{equation}
\label{sf06}4\sqrt{(kp)^2-m_e^2m_l^2}=4k_0(p_0+p_1)=4k_0p_0(1+v_1) 
\end{equation}
corresponds for some fixed $p$ to the subset of partons moving with velocity 
$\overrightarrow{v}=\overrightarrow{p}/p_0$. If this velocity has the
opposite direction to the probing electron, then after passing through the
whole subset $G(p_0)d^3p$ the electron has not still reached backward
boundary of the proton, where meanwhile the new partons appeared. And on
contrary, if the velocity of subset has the same direction as the electron,
then not all of these partons have the same chance to meet this electron.
Namely, the partons close to the backward boundary are excluded from the
game sooner than the electron reaches them. Quantitatively, the number of
partons limited by the proton volume and having chance to meet the electron
(with velocity $\sim 1)$ will be 
\begin{equation}
\label{sf07}dN=(1+v_1)G(p_0)d^3p 
\end{equation}
Including this correction to the flux factor (\ref{sf06}), then instead of
Eq.(\ref{sf3}) we get the tensor equation%
$$
P_\alpha P_\beta \frac{W_2}{M^2}-g_{\alpha \beta }W_1+A(P_\alpha q_\beta
+P_\beta q_\alpha )+Bq_\alpha q_\beta = 
$$

\begin{equation}
\label{sf5}=\frac{P_0}M\int \frac{G(p_0)}{p_0}(2p_\alpha p_\beta -g_{\alpha
\beta }\,pq)\delta ((p+q)^2-m^2)d^3p,\qquad p_0=\sqrt{m^2+p_1^2+p_2^2+p_3^2} 
\end{equation}
for which (\ref{sf2}){\it \ is not required}. The terms with the functions $%
A $ and $B$ do not contribute to the cross section (since $q_\alpha
K^{\alpha \beta }=q_\beta K^{\alpha \beta }=0$), but generally must be
included to ensure the equation consistence if the tensors are not gauge
invariant. Also let us note, the correction similar to (\ref{sf07}) was not
used in the Eq.(\ref{sf3}) since due to (\ref{sf2}) the all partons in the
applied approach have the same velocity as the proton.

Now the contracting of (\ref{sf5}) with tensors $g^{\alpha \beta },q^\alpha
q^\beta ,P^\alpha P^\beta ,P^\alpha q^\beta $ gives in the result set of
four equations 
\begin{equation}
\label{r1}W_2-4W_1+2M\nu (A-xB)=\frac 1{M\nu }\int \frac{G(p_0)}{p_0}%
[m^2-2Mx\nu ]\delta \left( \frac{pq}{M\nu }-x\right) d^3p 
\end{equation}
\begin{equation}
\label{r2}\frac \nu {2Mx}W_2+W_1-2M\nu (A-xB)=\frac 1{M\nu }\int \frac{%
G(p_0) }{p_0}[Mx\nu ]\delta \left( \frac{pq}{M\nu }-x\right) d^3p 
\end{equation}
\begin{equation}
\label{r3}W_2-W_1+\nu (2MA+\nu B)=\frac 1{M\nu }\int \frac{G(p_0)}{p_0}%
[p_0^2-\frac{Mx\nu }2]\delta \left( \frac{pq}{M\nu }-x\right) d^3p 
\end{equation}
\begin{equation}
\label{r4}W_2-W_1+(M\nu -2M^2x)A-2Mx\nu B=\frac 1{M\nu }\int \frac{G(p_0)}{%
p_0}[p_0Mx-\frac{Mx\nu }2]\delta \left( \frac{pq}{M\nu }-x\right) d^3p 
\end{equation}
in which the $\delta -$function from the integral (\ref{sf5}) is expressed 
\begin{equation}
\label{sf6}\delta ((p+q)^2-m^2)=\delta (2pq+q^2)=\delta \left( 2M\nu \left( 
\frac{pq}{M\nu }-\frac{Q^2}{2M\nu }\right) \right) =\frac 1{2M\nu }\delta
\left( \frac{pq}{M\nu }-x\right) 
\end{equation}
If we define 
\begin{equation}
\label{sf11}V_j(x)\equiv \int G(p_0)\left( \frac{p_0}M\right) ^j\delta
\left( \frac{pq}{M\nu }-x\right) d^3p,\qquad j=-1,0,1 
\end{equation}
then the solution of the set (\ref{r1})-(\ref{r4}) reads 
\begin{equation}
\label{r5}
\begin{split}
2MW_1&=\frac \nu {2Mx+\nu }\\&\quad\cdot \left\{ V_{-1}(x)\left[ x-\frac M\nu
\left( \frac{m^2}{M^2}-x^2\right) -2\frac{m^2x}{\nu ^2}\right] +V_0(x)\frac{%
2Mx}\nu +V_1(x)\frac{2M^2x}{\nu ^2}\right\} 
\end{split}
\end{equation}
\begin{equation}
\label{r6}
\begin{split}
\nu W_2&=x\left( \frac \nu {2Mx+\nu }\right) ^2\\&\quad\cdot \left\{
V_{-1}(x)\left[ x-\frac M\nu \left( \frac{m^2}{M^2}+x^2\right) -2\frac{m^2x}{%
\nu ^2}\right] +V_0(x)\frac{6Mx}\nu +V_1(x)\frac{6M^2x}{\nu ^2}\right\} 
\end{split}
\end{equation}
\begin{equation}
\label{r7}
\begin{split}
\nu ^2MA&=-\left( \frac \nu {2Mx+\nu }\right) ^2\\&\quad\cdot \left\{
V_{-1}(x)\left[ \frac 12\left( \frac{m^2}{M^2}+3x^2\right) +\frac{m^2x}{M\nu 
}\right] -V_0(x)2x\left( 1-\frac{Mx}\nu \right) -V_1(x)\frac{3Mx}\nu
\right\} 
\end{split}
\end{equation}
\begin{equation}
\label{r8}
\begin{split}
\nu ^3B&=\left( \frac \nu {2Mx+\nu }\right) ^2\\&\quad\cdot \left\{
V_{-1}(x)\left[ \frac 12\left( \frac{m^2}{M^2}+3x^2\right) +\frac{m^2x}{M\nu 
}\right] -V_0(x)3x+V_1(x)\left( 1-\frac{Mx}\nu \right) \right\} 
\end{split}
\end{equation}
For next discussion we assume $\nu \gg M$, then 
\begin{equation}
\label{r9}\nu W_2\equiv F_2(x)=x^2V_{-1}(x),\qquad MW_1\equiv F_1(x)=\frac
x2V_{-1}(x) 
\end{equation}
so it is obvious the Callan-Gross relation $2xF_1=F_2$ holds in this
approximation.

In the next step, following the Eq.(\ref{kd10}), we accept the approximation

\begin{equation}
\label{sf7}\frac{pq}{M\nu }\approx \frac{p_0+p_1}M
\end{equation}
then the integrals (\ref{sf11}) can be expressed 
\begin{equation}
\label{r10}V_j(x)=\int G(p_0)\left( \frac{p_0}M\right) ^j\delta \left( \frac{%
p_0+p_1}M-x\right) d^3p,\qquad j=-1,0,1
\end{equation}
This relation with the use of (\ref{eq4}),(\ref{eq20}) implies 
\begin{equation}
\label{r11}\left( \frac{p_0}M\right) ^jP(p_0)=\mp \frac 2MV_j^{\prime
}(x_{\pm })x_{\pm },\qquad j=-1,0,1
\end{equation}
where $x_{\pm }$ is defined in (\ref{eq19b}). The relations (\ref{r11}) and (%
\ref{eq19c}) give 
\begin{equation}
\label{r12}\frac{V_j^{\prime }(x)}{V_k^{\prime }(x)}=\left( \frac{p_0}%
M\right) ^{j-k}=\left( \frac{x_{+}+x_{-}}2\right) ^{j-k}=\left( \frac x2+
\frac{x_0^2}{2x}\right) ^{j-k},\qquad x_0=\frac mM
\end{equation}
In the previous section we have shown such functions as (\ref{r10}) obey the
relation (\ref{eq24a}), which means in particular, that the functions have a
maximum at $x_0$ and vanish for $x\leq x_0^2$. Therefore the same statement
is valid also for functions $F_2/x^2$ and $F_1/x$ from Eq.(\ref{r9}) 
\begin{equation}
\label{r121}\frac{F_2(x_{+})}{x_{+}^2}=\frac{F_2(x_{-})}{x_{-}^2},\qquad 
\frac{F_1(x_{+})}{x_{+}}=\frac{F_1(x_{-})}{x_{-}}
\end{equation}
This means that the structure functions of our idealized hadron also have
the maximum at $x_0$ or higher, if the peak of $V_{-1}(x)$ is not rather
sharp. Obviously, the peak will be sharp if $P(p_0)\neq 0$ for $p_0=m$. At
the same time, it should be kept in mind, that due to (\ref{kc10}) any
function expressed in ''real'' variable $x_B$ will be slightly smeared in
view of this function expressed in $x$. That is just the case of the
integrals (\ref{sf11}) approximated by (\ref{r10}). But this smearing should
be quite negligible for very low $x_B$ and high $Q^2$, see end of section 2.

Further, our considerations have started in previous section from the
distribution function $F(x)$ for which we have obtained relation (\ref{eq20}%
). The combination of this equation with (\ref{r9}), (\ref{r11}) and (\ref
{eq19c}) gives 
\begin{equation}
\label{r122}P(p_0)=-\frac 1M\left( \frac{F_2(x)}{x^2}\right) ^{\prime
}(x^2+x_0^2),\qquad x=\frac{p_0+\sqrt{p_0^2-m^2}}M 
\end{equation}
\begin{equation}
\label{r13}F^{\prime }(x)=\frac 12\left( \frac{F_2(x)}{x^2}\right) ^{\prime
}\left( x+\frac{x_0^2}x\right) 
\end{equation}
How to compare the last equation with the standard relation (\ref{sf4}) for $%
F$ and $F_2$? As we have already told, the standard approach (\ref{sf3}) is
exact in the case when the partons are static with respect to the nucleon,
i.e. when $x=m/M$. The Eq.(\ref{sf5}) itself is more exact, but the further
procedure with it requires the masses of the all partons in the considered
subset being equal. Therefore for a comparison let us consider first the
extreme scenario when the parton distribution functions $F(x)$ and $P(p_0)$
are (see Eq.(\ref{eq20})) rather narrowly peaked around the points $x_0=m/M$
and $p_0=m$. Then for $x\approx x_0$ Eq.(\ref{r13}) gives 
\begin{equation}
\label{r14}F^{\prime }(x)=\frac 12\left( \frac{F_2(x)}{x^2}\right) ^{\prime
}\left( x+\frac{x_0^2}x\right) \simeq \frac 12\frac{F_2^{\prime }(x)}{x_0^2}%
(x_0+x_0)=\frac{F_2^{\prime }(x)}{x_0} 
\end{equation}
from which the second relation (\ref{sf4}) follows as a limiting case of (%
\ref{r13}) 
\begin{equation}
\label{sf33}x_0F(x_0)\approx F_2(x_0) 
\end{equation}
Now, in the realistic case when the distribution functions are broad, the
exact validity of (\ref{sf3}) again requires static partons, therefore the
corresponding distribution function represents also a spectrum of masses.
But then obviously the above procedure for a single $m$ can be repeated with
spectrum of masses $F(x_0)$ giving in the result instead of (\ref{sf33}) the
relation 
\begin{equation}
\label{r15}\int x_0F(x_0)\delta (x-x_0)dx_0=\int F_2(x_0)\delta (x-x_0)dx_0 
\end{equation}
which implies 
\begin{equation}
\label{r16}xF(x)=F_2(x) 
\end{equation}
In this sense the approach based on Eq.(\ref{sf3}) can be understood as a
limiting case of that based on Eq.(\ref{sf5}).

\subsection{The high order corrections}

The considerations of previous subsection are based on the approximation (%
\ref{sf7}) which in the result gives relations (\ref{r122}), (\ref{r13}).
Actually we had to calculate integrals (\ref{sf11}) instead of (\ref{r10})
differing in the argument of $\delta $-function according to Eq.(\ref{kd10}%
). The integrals (\ref{sf11}) cannot be solved analytically according to the
recipe for Eq.(\ref{eq4}), however in principle their solution can by
obtained by iterations. For example, first equation (\ref{r9}) reads 
\begin{equation}
\label{sf48}F_2(x)=x^2\int G(p_0)\frac M{p_0}\delta \left( \frac{pq}{M\nu }%
-x\right) d^3p 
\end{equation}
Let us have some $F_2$, then algorithm of iterative procedure could be
following:

0. step: $G_0$ is given by Eq.(\ref{r122}), $G$ is related to $P$ by (\ref
{eq18}).

1. step: Insert $G_0$ into (\ref{sf48}), result of integration is some
function $f_1(x).$ Make the difference $\Delta _1F_2=F_2-f_1$ and insert $%
\Delta _1F_2$ into (\ref{r122}), the relation gives the corresponding
correction $\Delta _1G$. The result of this iterative step is $%
G_1=G_0+\Delta _1G.$ Then next steps will follow by analogy, on the end the
corrected $G$ should be obtained.

More detailed discussion of considered correction exceeds scope of this
paper and requires further study. The correction should be rather small, but
let us remark that its evaluation requires some assumption about mass $m$
(or spectrum of masses). Also, let us note this correction together with
terms $O(1/\nu )$ in (\ref{r5}),(\ref{r6}) introduces into the structure
functions some $Q^2$ dependence having purely kinematic origin (we still
assume $G$ being $Q^2$ independent). Obviously, all these corrections vanish
for $Q^2\rightarrow \infty .$

\section{ Discussion}

\setcounter{equation}{0}

Are the considerations suggested in previous section compatible with the
assumptions and philosophy of QPM and all that, is it legally to speak about
distribution function in LAB? First, let us shortly recall standard
interpretation of DIS in framework of QPM.

In the classical experiment, e.g. BCDMS \cite{BCDMS} muons scatter on proton
target at rest in the {\it laboratory system}. From measured angles and
energies of the scattered muons one determines the invariant cross section
as the function of kinematic invariants $x_B,Q^2.$ Next, from this cross
section the electromagnetic structure function $F_2(x_B,Q^2)$ is evaluated.
The fact, that for sufficiently big $Q^2$ the structure function
(approximately) scales $F_2(x_B,Q^2)\approx F_2(x_B),$ leads to the
conclusion that in the experiment actually the scattering of two {\it %
point-like} particles takes place. This experimental fact is a basic
motivation of the QPM in which it is postulated that the nucleon contains
point-like electromagnetically active particles (partons), which can be for
sufficiently high $Q^2$ treated as effectively free and their interaction
with the muon is described by Feynman diagram with one photon exchange. That
also means the struck partons remain on-mass shell. These assumptions should
be fulfilled first of all in the system, where our experiment is done, i.e.
in LAB. Of course, another point is, that in this system the picture of
partons is in some respect obscured by the fact, that we do not know more
about the kinematics of partons, their momenta, energies. The picture is
quite clarified, when we change over from LAB to the IMF. Then the masses of
partons do not play any role and energy is the same as momentum.
Simultaneously, the invariant parameter $x_B$ obtains simple physical sense
- fraction of proton energy carried by the parton. And only now the quark -
parton distribution functions can be introduced and their known connection
with the structure function shown.

The difference between this standard approach and that of ours can be well
seen by comparing of Eqs.(\ref{sf3}), (\ref{sf5}). The general philosophy
according to which the scattering of charged lepton on a nucleon in DIS is
realized via scattering on point-like charged partons is common for both
equations. The actual difference is rather only technical consisting in the
choice of integration variables and approximations enabling to evaluate the
integrals.

The practical consequence of more simplifying approach based on Eq.(\ref{sf3}%
) is that resulting picture has good sense only in IMF where also problem of
parton masses is completely separated off, which can be even useful.

On the other hand, the approach based on Eq.(\ref{sf5}), requiring in
addition only assumption about the nucleon spherical symmetry, takes
consistently into account parton transversal momenta and is not confined to
some preferred system (even though our results are presented in LAB). There
is one important consequence, namely in this description the parton masses,
or more exactly ratio $m/M$ appeared as a free parameter.

\begin{figure}
\begin{center}
\epsfig{file=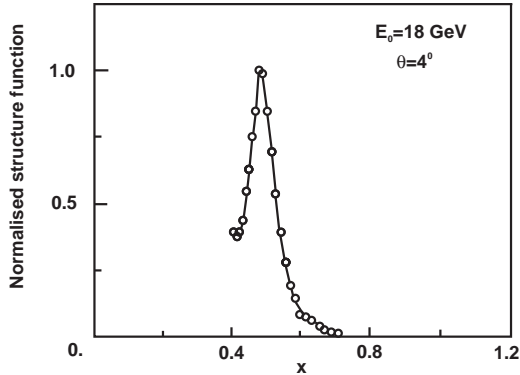, height=5cm}
\end{center}
\caption {The structure function for quasi-elastic $e^- d$ scattering, 
see text}
\label{fg4}
\end{figure}

Any speculation about parton mass already goes beyond postulates of QPM,
nevertheless look on some experimental data. Before coming to the proton
structure function, let us look at the Fig.\ref{fg4}, 
where the ''structure function'' of the deuteron measured in quasi-elastic $%
e^{-}d$ scattering \cite{D2} is shown, clearly proving the presence of two
nucleons in the nucleus. The similarity with general picture Fig.\ref{fg2}
is well seen. The kinematics of the two nucleons in the deuteron rest system
implies

\begin{equation}
\label{d3}{\it \sqrt{m^2+\left| \overrightarrow{p_1}\right| ^2}+\sqrt{%
m^2+\left| \overrightarrow{p_2}\right| ^2}=M_D,\qquad \overrightarrow{p_1}=- 
\overrightarrow{p_2}} 
\end{equation}
where $m$ should be understood as some {\it effective mass} which, due to
binding is slightly less then $M_D/2$. This difference roughly corresponds
to the depth of the potential if non-relativistic approach is used. From (%
\ref{d3}) the kinematically allowed region for corresponding $x$ easily
follows

\begin{equation}
\label{d4}0.5-\Delta x\leq x\leq 0.5+\Delta x,\qquad \Delta x\equiv \frac 12 
\sqrt{1-\left( \frac{2m}{M_D}\right) ^2} 
\end{equation}

In the case of partons inside the nucleon the situation is much more
delicate. The interaction among the quarks and gluons is very strong,
partons themselves are mostly in some shortly living virtual state, is it
possible to speak about their mass at all? Strictly speaking probably not.
The mass in exact sense is well defined only for free particles, whereas the
partons are never free by definition. Therefore let us try to speak at least
about an {\it effective mass}. By this term we roughly mean the mass that a
free parton would have to have to interact with the probing lepton equally
as our bounded one. Intuitively, this mass should correlate to $Q^2$. A
lower $Q^2$ allows more time and space for struck parton to interact with
some others, in the result the energy is transferred to a greater system
than the parton itself. On the contrary, the higher $Q^2$ should mediate
interaction with more ''isolated'' parton.

\begin{figure}
\begin{center}
\epsfig{file=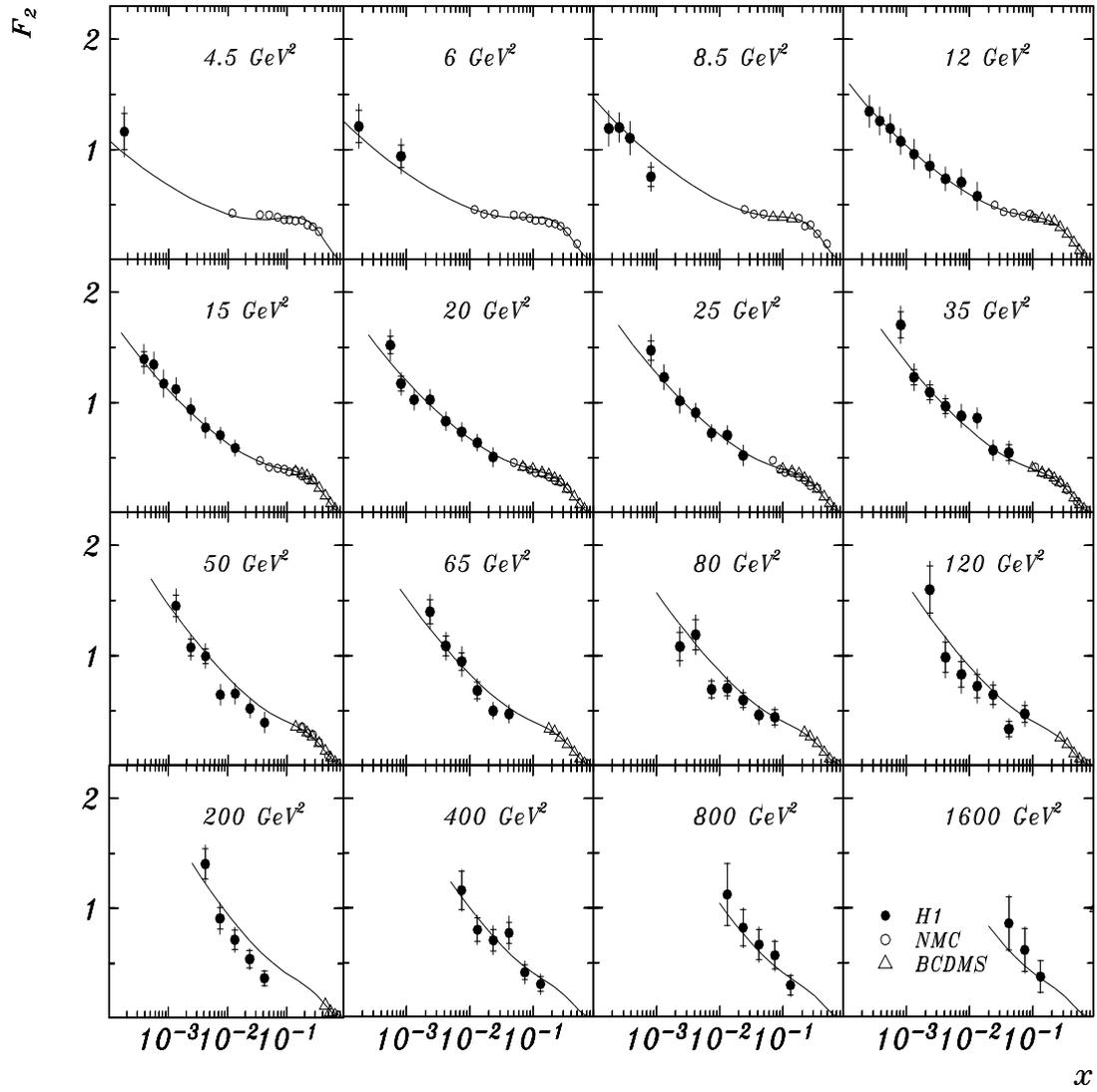, width=16.5cm}
\end{center}
\caption{The structure function $F_2 (x,Q^2 )$ taken from [12].}%
\label{fg5}
\end{figure}

Now let us try the formulae from previous section (with suggested sense of
mass $m$) confront with the experimental data. In the Fig.\ref{fg5} recently
obtained picture of the proton structure function $F_2$ \cite{h1f2} is
shown. 
No peak of that sort in Fig.\ref{fg2} or Fig.\ref{fg4} according to Eq.(\ref
{r121}) is seen. There are two extreme alternatives.

1) The effective mass $m/M$ of quarks can be for given $Q^2$ well
represented by one number. Then, obviously this value should be below the
experimental limit of $x$ $\approx 10^{-3}\div 10^{-4}.$

2) The concept of effective mass reflects even for fixed $Q^2$ rather some
distribution than a single value. Then the structure function is some
superposition of curves similar to that in Fig.\ref{fg2} , but with
different positions of their maxima. Such superposition could be generated
not only by different flavors but also by the components commonly
denoted by the term valence and sea quarks with distribution functions given
e.g. in \cite{parval}. We shall not discuss the scenario of effective mass
distribution in general, but only check one extreme: the case of static
partons mentioned just below Eq.(\ref{sf4}). These partons exactly obey the
equation $p_\alpha =xP_\alpha $. Obviously having measured $F_2$ from the
limit $x\geq L$, one can estimate the mean value 
\begin{equation}
\label{r17}\frac{<m>}M<\frac{\int_L^1xF(x)dx}{\int_L^1F(x)dx}=\frac{%
\int_L^1F_2(x)dx}{\int_L^1\frac{F_2(x)}xdx}
\end{equation}
The numerical calculation with the function fitting the data in the Fig.\ref
{fg5} 
\begin{equation}
\label{d5}F_2(x,Q^2)=[3.07x^{0.75}+0.14x^{-0.19}(1-2.93\sqrt{x})(\ln
Q^2-0.05\ln {}^2Q^2)](1-x)^{3.65}
\end{equation}
gives the value $<m>$ in the region of tens $MeV$ depending on $L$ and $Q^2$
very roughly as
\begin{equation}
\label{r18}\frac{<m>}M<1.8\frac{L^{0.34}}{\ln Q^2},\qquad 10^{-4}\leq L\leq
10^{-2},\qquad 20\leq Q^2\leq 1600GeV^2
\end{equation}
Obviously, this scenario is less restrictive than the first one. 

It is possible, that the real case is somewhere between the two mentioned
extremes. At the same time, the $Q^2$ dependence in Fig.\ref{fg5} could be
qualitatively understood in the manner suggested above: the higher $Q^2$
prefers to mediate interactions with partons having less effective mass,
therefore for higher $Q^2$ the low $x$ region should be more populated.
Apparently, the quantitative expression of this correspondence is problem of
dynamics.

\section{Summary}

In the present paper we have discussed a connection between the parton
distribution functions ordinarily defined in the infinite momentum frame and
the analogous functions defined in the hadron rest system. Assuming
spherical symmetry of the hadron and an equal effective mass $m$ of the all
partons of considered sort we have shown:

1) There exists unambiguous relation between the distribution functions
defined in the both reference systems.

2) The proposed approach taking consistently parton transversal momenta into
account gives the relation between the (electromagnetic) structure and
distribution function somewhat modified in regard of the standard one.
However, the standard relation is involved in that of our as a limiting
case. The  approach is not connected to any preferred reference system and
explicitly involves ratio $m/M$ as a free parameter.

3) Within our approach in the structure functions we have identified  some
rather small,  $Q^2$-dependent terms having purely kinematic origin. 

4) The resulting relations pose the constraint on the shape of structure and
distribution functions, which implies in particular the functions have the
maximum at $x\approx m/M$ and vanish for $x<m^2/M^2.$

5) We compared our results with the data on proton structure function $(F_2)$
assuming the two rather extreme scenarios:

i) The effective mass is for a fixed $Q^2$ well represented by one number,
then the ratio $m/M$ is below presently reached limit of $x$ $(10^{-3}\div
10^{-4}).$

ii) The effective mass is at given $Q^2$ represented by some distribution
and moreover the partons are static. Then the present data suggest the value 
$<m>/M$ should be at most of order $10^{-2}$.

Simultaneously, the $Q^2$ -dependence of structure function is qualitatively
interpreted as a result of dynamic correlation of the effective mass and $Q^2
$.
\\ \\ {\bf Acknowledgements} I would like to express my gratitude to J.Pi\v
s\'ut for many inspiring discussions which in the result motivated this
work. I am also indebted to J.Ch\'yla for critical reading of the manuscript
and valuable comments.

\end{document}